\documentclass[conference]{IEEEtran}
\usepackage{myStyleIEEE}
\usepackage{calc}
\usepackage{ushort}
\IEEEoverridecommandlockouts

\usepackage{float}

\usepackage{siunitx}
\AtBeginDocument{\DeclareSIUnit{\MWh}{MWh}}
\AtBeginDocument{\DeclareSIUnit{\kWh}{kWh}}
\AtBeginDocument{\DeclareSIUnit{\voltampere}{VA}}

\usetikzlibrary{spy}
\usepackage{amsmath}
\usepackage{mathtools}
\usepackage{isomath}

\usepackage{eurosym}
\DeclareRobustCommand{\officialeuro}{%
  \ifmmode\expandafter\text\fi
  {\fontencoding{U}\fontfamily{eurosym}\selectfont e}}

\def\BibTeX{{\rm B\kern-.05em{\sc i\kern-.025em b}\kern-.08em
    T\kern-.1667em\lower.7ex\hbox{E}\kern-.125emX}}

\usepackage{etoolbox}
\usepackage{letltxmacro}

\usepackage{acronym}
\usepackage[]{optidef}
\usepgfplotslibrary{groupplots}
\definecolor{darkgreen}{rgb}{0.0, 0.5, 0.0}
\pgfplotsset{
    mark first last/.style={
        scatter,
        scatter src=x,
        scatter/@pre marker code/.code={%
            \pgfmathtruncatemacro\isfirst{\coordindex==0}%
            \pgfmathtruncatemacro\islast{\coordindex==\numcoords-1}%
            \ifnum\isfirst=1
                \pgfplotsset{mark=asterisk, mark options={solid}} % Solid star for first
            \else
                \ifnum\islast=1
                    \pgfplotsset{mark=o, mark options={solid}} % Solid circle for last
                \else
                    \pgfplotsset{mark=none}% No marker for others
                \fi
            \fi
        },
        scatter/@post marker code/.code={}
    }
}

\begin{document}

\title{Revisiting Power System Stabilizers with Increased Inverter-Based Generation: A Case Study}

\author{\IEEEauthorblockN{Jovan Krajacic}
\IEEEauthorblockA{\textit{Power Systems Laboratory} \\
ETH Z{\"u}rich, Switzerland \\
jkrajacic@ethz.ch}
\and
\IEEEauthorblockN{Keith Moffat}
\IEEEauthorblockA{\textit{Automatic Control Laboratory} \\
ETH Z{\"u}rich, Switzerland \\
kmoffat@ethz.ch}
\and
\IEEEauthorblockN{Gustavo Valverde}
\IEEEauthorblockA{\textit{Power Systems Laboratory} \\
ETH Z{\"u}rich, Switzerland \\
gustavov@ethz.ch}
\thanks{Research partially supported by NCCR Automation, a National Centre of Competence in Research, funded by the Swiss National Science Foundation (grant number 51NF40\_180545).}}

\maketitle
\IEEEpeerreviewmaketitle
% \IEEEpubidadjcol

%ABSTRACT
\begin{abstract}
As power systems evolve with increasing production from Inverter-Based Resources (IBRs), their underlying dynamics are undergoing significant changes that can jeopardize system operation, leading to poorly damped oscillations or small-signal rotor angle instability.  In this work, we investigate whether Power System Stabilizer (PSS) setting adjustments can effectively restore system stability and provide adequate damping in systems with increased IBR penetration, using the benchmark Kundur Two-Area System as a case study. Specifically, we evaluate the model-based Residues and P-Vref PSS tuning methods to examine their effectiveness under evolving grid conditions. Our findings indicate that the effectiveness of these tuning methods is not guaranteed, particularly when coordination is limited. Consequently, our case study motivates local and adaptive online PSS tuning methods.
\end{abstract}

%INDEX TERMS

%%%%%%%%%%%%%%%%%%%%%%%%%%%%%%%%%%%%%%%%%%%%%%%%%%%%%%%%%%%%%%%%%%%%%%%%%%%

\section{Introduction} \label{sec:introduction}

% P1 - Describe the background of the problem $\rightarrow$ integration of renewables in power systems $\rightarrow$ decommissioning of SGs $\rightarrow$ oscillations

As power systems continue to integrate renewable energy resources, conventional Synchronous Generators (SGs) are progressively being replaced by Inverter-Based Resources (IBRs). This transition alters grid dynamics by removing mechanical inertia and damping \cite{gfl}, potentially  jeopardizing small-signal stability, defined as the ability of a power system to maintain synchronism under small disturbances \cite{Kundur1994}.

% P2 - Talk about the oscillations - how and where do they appear, what are the types, etc. Talk about the literature that reviews them

Several studies have examined how increased IBR penetration affects the electromechanical modes inherent to conventional power systems \cite{cse_interaction, 9110789, epri_gfm, en15124273}. Specifically, the authors of \cite{cse_interaction} and \cite{9110789} discuss the impact that IBRs have on the existing modes in the system, highlighting that weak interconnections between IBRs and the grid, combined with decreasing system inertia and interactions between newly added and existing IBRs, can have a detrimental effect on system stability. Furthermore, \cite{epri_gfm} and \cite{en15124273} compared the converter-interfaced generation control strategies and examined their impact on grids with increasing penetration of IBRs, concluding that future power systems with a high share of non-synchronous generation may experience a reduction in small-signal stability, depending on the IBR control strategy. 
% In addition to these works, \cite{6848832} points to potential instabilities rising from the positive feedback induced by the phase-locked loop of the Grid Following Converters (GFLs) in weak grids, leading to a destabilizing control response. % that reinforces the disturbance rather than damping it.

% P3 - Many address the IBRs and not the PSSs; only a few focus on the PSSs. 

While most research has focused on the occurrence of oscillations and the role IBRs play in their emergence, comparatively few studies have explored the performance of existing Power System Stabilizers (PSSs) in grids with high IBR penetration, originally designed to mitigate electromechanical oscillations in SG-dominated systems \cite{Kundur1994}. The work in \cite{cse_interaction}, \cite{en15124273}, and \cite{beyond_low_inertia} emphasized the need to re-tune PSS parameters since they were tuned decades ago for grids with dynamics dominated by SGs. Furthermore, the authors in \cite{uros_small_signal} examined the transition from SG-dominated grids to IBR-dominated systems, including Grid Following (GFL) converters, highlighting potential instabilities arising from the interactions between GFL converters, PSSs, and Automatic Voltage Regulators (AVRs). Similar observations were made in \cite{stability_assesment}, which identified PSSs and AVRs as potential sources of instability in systems with high IBR penetration.  

% P4 - Contributions of this work
% Building on the findings of previous work, 
Using the Kundur Two-Area Test System as a benchmark case study, we investigate the effectiveness of legacy-tuned PSSs when a subset of the SGs are replaced by GFL converters. We use GFL inverter control in our case study as it is the prevailing control strategy today \cite{dorfler_dominic_low_inertia}. Subsequently, we investigate the efficacy of re-tuning the PSSs to restore system stability. Specifically, we investigate both ``coordinated'' and ``local'' PSS tuning approaches, in which communication between SGs is and is not allowed, respectively. We evaluate the performance of industry-standard, model-based PSS tuning methods (Residues and P-Vref) and identify additional requirements---specifically, coordination---needed for these methods to be effective. 

Based on our tests on the benchmark Kundur Two-Area Test System, we draw the following findings. The PSSs remaining in the system after replacing SGs with GFL converters require re-tuning, as their initial settings may lead to system instability. When re-tuning just one PSS, the Residues method outperforms the P-Vref method; however, the performance can still be unsatisfactory. In contrast, when both PSSs are re-tuned, the P-Vref method outperforms the Residues method, which fails to stabilize the system when coordination is unavailable. Although re-tuning both PSSs with P-Vref stabilizes the test cases, it is not guaranteed to work for all systems.

% P5 - Paper Organization
The remainder of this paper is organized as follows. Section~\ref{sec:preliminaries} provides an overview of small-signal stability and PSSs. Section~\ref{sec:methodology} reviews the Residues and P-Vref methods. Section~\ref{sec:results} presents the case study and summarizes the main findings, while Section~\ref{sec:conclusion} concludes the paper.

%%%%%%%%%%%%%%%%%%%%%%%%%%%%%%%%%%%%%%%%%%%%%%%%%%%%%%%%%%%%%%%%%%%%%%%%%%%
\section{Preliminaries} \label{sec:preliminaries}

\subsection{System Definition} \label{preliminaries:small_signal}
Let us describe the power system model using a set of $n$ first-order differential equations and $m$ algebraic equations:
\begin{subequations}
\begin{align}
    \dot{\mathbf{x}} &= \mathbf{f}(\mathbf{x}, \mathbf{y}, \mathbf{u}) \,,\label{eq:system_x}\\
    \mathbf{0} &= \mathbf{g}(\mathbf{x}, \mathbf{y}, \mathbf{u}) \,,\label{eq:system_0} \\
    \mathbf{z} &= \mathbf{h}(\mathbf{x}, \mathbf{y}, \mathbf{u})\,,\label{eq:system_z}
\end{align}
\end{subequations}
where $\mathbf{x} \in \mathbb{R}^n$ denotes the vector of state variables, $\mathbf{y} \in \mathbb{R}^m$ the vector of algebraic variables, $\mathbf{u} \in \mathbb{R}^r$ the vector of inputs, and $\mathbf{z} \in \mathbb{R}^s$ the vector of outputs.

We assume that the system operates at a specific steady-state operating point at which a perturbation is introduced. Since the perturbation is assumed to be small, the non-linear functions $\mathbf{f}$, $\mathbf{g}$, and $\mathbf{h}$ can be approximated with a first-order Taylor Series expansion for $\Delta \mathbf{x}$, $\Delta \mathbf{y}$ and $\Delta \mathbf{z}$, the difference between $x$, $y$, and $z$ from their respective steady-state quantities. This gives
\begin{subequations}
\begin{align}
    \Delta \dot{\mathbf{x}} &= 
    \mathbf{A}
    \Delta \mathbf{x} +
    \mathbf{B}
    \Delta \mathbf{u}\,,  \label{eq:mat_ab} \\
    \Delta \mathbf{z} &= 
    \mathbf{C}
    \Delta \mathbf{x} +
    \mathbf{D}
    \Delta \mathbf{u}\, , \label{eq:mat_cd}
\end{align}
\end{subequations}
where $\mathbf{A} \in \mathbb{R}^{n \times n}$ is the state matrix, $\mathbf{B} \in \mathbb{R}^{n \times r}$ is the input matrix, $\mathbf{C} \in \mathbb{R}^{s \times n}$ is the output matrix, and $\mathbf{D} \in \mathbb{R}^{s \times r}$ is the feedthrough matrix, which is usually zero in electrical power systems. The system is asymptotically stable if the real parts of $\mathbf{A}$'s eigenvalues ($\lambda_i$ for $i = 1, \dots, n$) are negative.

\subsection{Power System Stabilizers} \label{preliminaries:pss}
To improve the small-signal stability of multi-machine power systems, it is common practice to integrate a PSS control loop into the generator's AVR system, as shown in Fig.~\ref{fig:block_diagram}. The PSS adds $V_\text{PSS}$ to $V_\text{error}$ to compensate for $V_\text{error}$ oscillations and provides a damping-torque component in phase with the rotor speed deviation $\Delta \omega$. For more details on the PSS structure, we refer the reader to \cite[Section~12.5]{Kundur1994}. % and \cite[Chapter~5]{small_signal}.

% The PSS input $\Delta \omega$ is filtered by a high-pass filter with time constant $T_W$ to remove speed variations unrelated to electromechanical oscillations, followed by lead-lag blocks for phase compensation. The output $V_\text{PSS}$ is limited to avoid AVR interference during large disturbances, while $K_\text{PSS}$ sets the damping contribution.

\subsection{Plant Transfer Function} \label{preliminaries:observability}
\begin{figure}[t!]
    \centering
    \scalebox{0.475}{\begin{tikzpicture}[%
    > = stealth, % arrow tips
    node distance=1.2cm and 1.5cm, % spacing between nodes
    box/.style={rectangle, draw, thick, minimum height=1.0cm, text width=2.5cm, align=center},
    sum/.style={circle, draw, thick, minimum size=0.5cm, inner sep=0pt, align=center},
    ]
    \node [sum] (sum1) {};
    \node [left of=sum1, node distance=1.25cm] (input1) {\Large $V_\text{ref}$};
    \draw [->, thick] (input1) -- (sum1) node[pos=0.85, yshift=1pt, below]{\large $+$};

    \node [sum, right of=sum1, node distance=2.0cm] (sum2){};
    \draw [->, thick] (sum1) -- (sum2) node[pos=0.95, yshift=1pt, below]{\large $+$} node[above, pos=0.45] (v1) {\Large $V_\text{error}$};

    \node[box, right=of sum2,  fill=gray!10] (regulator) {\large \textbf{AVR}};
    \node[box, right=0.5cm of regulator,  fill=gray!10] (exciter) {\large \textbf{Exciter}};
    \node[box, right=1.75cm of exciter,  fill=gray!10] (generator) {\large \textbf{Generator}};
    \draw [->, thick] (sum2) -- (regulator);
    \draw [->, thick] (regulator) -- (exciter);
    \draw [->, thick] (exciter) -- (generator);
    \node[above right=-0.55cm and 0.3cm of exciter] (efd) {\Large $E_{FD}$};
    
    \draw [->, thick] ($(generator.north east)+(0,-0.3)$) -- ++(2.5cm, 0);
    \draw [->, thick] ($(generator.south east)+(0,0.3)$) -- ++(2.5cm, 0);

    \draw [->, thick] ($(generator.south east)+(1.9,0.3)$) -- ++(0, 5.2cm) -| (sum1) node[pos=0.985, xshift=1pt, left]{\large $-$} ;
    \draw [->, thick] ($(generator.north east)+(1.3,-0.3)$) -- ++(0, 2.625cm) -- ++(-1.2, 0);

    % KPSS
    \node[box, above right=1.80 and -1.45 of generator, minimum height=1.2cm, text width=1.2cm] (kpss) {\large $K_\text{PSS}$};
    \node[above=0.08cm of kpss] {Gain};

     % Tw
    \node[box, left=0.70cm of kpss, minimum height=1.2cm, text width=1.5cm] (tw) {\large $\displaystyle \frac{sT_\mathrm{W}}{1+sT_\mathrm{W}}$};
    \draw [->, thick] (kpss) -- (tw);
    \node[align=center, above=0.00cm of tw] {High-pass Filter};

    % Leadlag1
    \node[box, left=0.70cm of tw, minimum height=1.2cm, text width=1.4cm] (leadlag1) {\large $\displaystyle \frac{1+sT_1}{1+sT_2}$};
    \draw [->, thick] (tw) -- (leadlag1);

    % Leadlag2
    \node[box, left=0.5cm of leadlag1, minimum height=1.2cm, text width=1.4cm] (leadlag2) {\large $\displaystyle \frac{1+sT_3}{1+sT_4}$};
    \draw [->, thick] (leadlag1) -- node[midway, above=0.6cm, align=center] {Lead/Lag Blocks} (leadlag2);

    % Output VPSS
    \node [left=4.0cm of leadlag2] (output1) {};
    \draw [->, thick] (leadlag2) -| (sum2) node[pos=0.975, xshift=1pt, left]{\large $+$};
    \node [left=2.8cm of leadlag2, yshift=8pt] {\Large $V_\text{PSS}$};

    % Limiter Block
    \draw [thick] ($(leadlag2.east)+(-2.3,-0.25)$) -- ++(-0.4,0) node[below, pos=0.5] {\large $\min$} -- ++(-0.4,0.5) coordinate[midway, shift={(-0.75,0)}] (midpoint) -- ++(-0.4,0) node[above, pos=0.5] {\large $\max$};
    \node[above left=0.05cm and 0.65cm of leadlag2, align=center] {Limiter};

    % Dashed block around Regulator and Exciter
    \node[draw=black, dashed, inner sep=0.4cm, fit={(regulator) (exciter) (generator)}] (dashedblock2) {};
    \node[draw=black, dashed, inner sep=0.4cm, inner ysep=0.6cm, fit={(kpss) (tw) (leadlag1) (leadlag2) (midpoint)}] (dashedblock3) {};
    \node[anchor=south] at (dashedblock2.north) {\large Plant Transfer Function $G(s)$};
    \node[anchor=south] at (dashedblock3.north) {\large PSS Transfer Function $F(s)$};

    \node[above right=0.2cm and 0.0cm of input1] (vc) {\Large $V_C$};
    \node [] at ($(kpss.east)+(1.0,0.25)$) {\Large $\Delta \omega$};

    \end{tikzpicture}}
    \vspace{-0.5cm}
    \caption{Block diagram of the Excitation system (AVR) and PSS control loops.}
    \label{fig:block_diagram}
    \vspace{-0.3cm}
\end{figure}
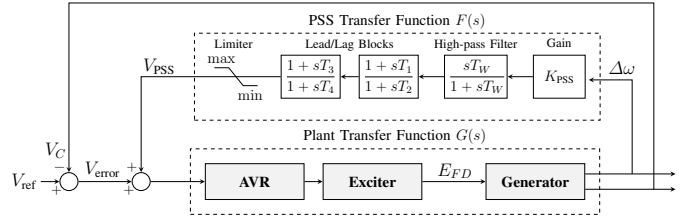

Let us now consider a power system model defined by \eqref{eq:mat_ab} and \eqref{eq:mat_cd}. We assume that a PSS is not connected and that the system has a single input $u$, corresponding to the AVR's reference voltage $V_\text{ref}$, and a single output $z$, which represents the rotor speed $\omega$ of the analyzed SG, as depicted in Fig.~\ref{fig:block_diagram}. We transform the state vector $\mathbf{x}$ using the left eigenvector matrix $\mathbf{W= \begin{bmatrix} \mathbf{w}_\mathrm{1}^\top &  \mathbf{w}_\mathrm{2}^\top & \dots & \mathbf{w}_\mathrm{n}^\top \end{bmatrix}^\top}$, where $\mathbf{w}_i \mathbf{A} =\lambda_i \mathbf{w}_i,$ and obtain $\Tilde{\mathbf{x}}=\mathbf{W}\mathbf{x}$. Substituting $\Tilde{\mathbf{x}}$ into \eqref{eq:mat_ab} and \eqref{eq:mat_cd} yields
\begin{subequations} \label{eq:transformed_system}
\begin{align}
    \dot{\tilde{\mathbf{x}}} &= \mathbf{W} \mathbf{A} \mathbf{W}^{-1} \tilde{\mathbf{x}} + \mathbf{W} \mathbf{b} u = \mathbf{\Lambda} \tilde{\mathbf{x}} + \mathbf{W} \mathbf{b} u \, ,\\
    z &= \mathbf{c} \mathbf{W}^{-1} \tilde{\mathbf{x}} + d u = \mathbf{c} \mathbf{V} \tilde{\mathbf{x}} + d u \, .
\end{align}
\end{subequations}
Presuming that $\mathbf{A}$ is diagonalizable, $\mathbf{\Lambda}$ is a diagonal matrix of eigenvalues.
Furthermore, $\mathbf{w}_i \mathbf{b}$ is a measure of the controllability of mode $i$, while $\mathbf{c}\mathbf{v}_i$ is a measure of the observability of mode $i$, where $\mathbf{V= \begin{bmatrix} \mathbf{v}_\mathrm{1} &  \mathbf{v}_\mathrm{2} & \dots & \mathbf{v}_\mathrm{n} \end{bmatrix}=W^{-1}}$.

The Transfer Function (TF) of the plant $G(s)$ in Fig.~\ref{fig:block_diagram} can be derived by applying the Laplace transform to \eqref{eq:transformed_system}:
\begin{equation}
\label{eq:G_s}
    G(s) = \frac{z(s)}{u(s)} \stackrel{d=0}{=} \sum_{i=1}^n \frac{R_i}{s - \lambda_i} \, ,
\end{equation} 
where the poles of $G(s)$ correspond to the eigenvalues of the state matrix $\mathbf{A}$ and $R_i=\mathbf{c} \mathbf{v}_i \mathbf{w}_i \mathbf{b}$ represents the residue of $G(s)$ associated with pole $\lambda_i$, which depends on both the observability and controllability of the mode.

%%%%%%%%%%%%%%%%%%%%%%%%%%%%%%%%%%%%%%%%%%%%%%%%%%%%%%%%%%%%%%%%%%%%%%%%%%%
%%%%%%%%%%%%%%%%%%%%%%%%%%%%%%%%%%%%%%%%%%%%%%%%%%%%%%%%%%%%%%%%%%%%%%%%%%%

%%%%%%%%%%%%%%%%%%%%%%%%%%%%%%%%%%%%%%%%%%%%%%%%%%%%%%%%%%%%%%%%%%%%%%%%%%%
\section{Methodology} \label{sec:methodology}
Our analysis focuses on two tuning approaches: the Residues and P-Vref methods, both of which are model-based, as they require a power system model for tuning the PSSs. We refer the reader to \cite[Section 6.2~and~5.8]{small_signal} for further details on the Residues and P-Vref methods, respectively.

\subsection{Residues Method} \label{methodology:resisudes}
The closed-loop TF of the system can be written in terms of $G(s)$ and the PSS's TF $F(s)$ depicted in Fig.~\ref{fig:block_diagram}:
\begin{equation}
    W(s)=\frac{\Delta \omega}{\Delta V_{\text{ref}}}=\frac{G(s)}{1-G(s)F(s)} \, . \label{eq:residues_tf}
\end{equation}
The residues method targets a specific critical closed-loop pole $\lambda_c=\sigma_c + j \omega_c$ of $W(s)$ such that the root-locus branch moves from $\lambda_c$ towards the left half of the complex plane for $K_\text{PSS}>0$. This movement is achieved by the phase compensation provided by the PSS lead-lag blocks $\theta_{H_\text{PSS}}$ at the unstable mode frequency $\omega_c$:
\begin{equation}
    \theta_{H_\text{PSS}}(j\omega_c) = \left| \theta_{R_c} - 180^\circ \right| \, ,
\end{equation}
where $R_c=r_c\angle{\theta_{R_c}}, \, \theta_{R_c} \in [0^\circ, \, 360^\circ] $. Assuming the PSS lead-lag blocks are equal, such that $T_1=T_3=T$ and $T_2=T_4=aT$, the maximum phase angle $\phi_m$ of a lead-lag block, which occurs at frequency $\omega_m=\frac{1}{T\sqrt{a}}$, equals
\begin{equation}
    \sin(\phi_m)=\frac{1-a}{1+a}\, . \label{eq:residues_max_angle}    
\end{equation}

Using \eqref{eq:residues_max_angle} and assuming the maximum phase compensation provided by each lead-lag block is approximately $\pm60^\circ$, the required number of lead-lag blocks equals $N=1$ if $|\theta_{H_\text{PSS}}| < 60^\circ$, $N=2$ if $60^\circ \leq  |\theta_{H_\text{PSS}}|$ and $N=3$ if $120^\circ \leq  |\theta_{H_\text{PSS}}|$. The parameters of each lead-lag block are then given as:
\begin{equation}
    a = \frac{1 - \sin\left(\frac{\theta_{H_\text{PSS}}}{N}\right)}{1 + \sin\left(\frac{\theta_{H_\text{PSS}}}{N}\right)} \,, \quad 
    T = \frac{1}{\omega_c \sqrt{a}} \,.
\end{equation}

Typically, the value of $T_\mathrm{W}$ in the high-pass filter ranges from 1 to 20 s, depending on the target frequency range. Furthermore, the PSS gain $K_\text{PSS}$ may be determined by evaluating the damping provided to the mode $\lambda_c$ using the root locus of the TF \eqref{eq:residues_tf}.

\subsection{P-Vref Method} \label{methodology:pvref}

The P-Vref method uses a different TF, $G_{P\text{-}V_{\text{ref}}}(s)$, to tune the PSS parameters. This TF maps the reference voltage $V_{\text{ref}}$ to the generator's active power output $P$ and is computed with the shaft dynamics disabled, i.e., $\Delta \omega = 0$. 
After connecting $F(s)$, the PSS's TF which maps $\Delta \omega$ to $V_\text{PSS}$, the compensated plant TF $G_C(s)$ can be written as \cite[Section~5.8]{small_signal}:
\begin{equation}
    G_C(s)=\frac{\Delta P}{\Delta \omega}= G_{P\text{-}V\text{ref}} (s) F(s) \, .
\end{equation}

As stated in Section~\ref{preliminaries:pss}, the purpose of the PSS is to provide a damping-torque component $T_e$ in phase with the rotor speed deviation $\Delta \omega$. 
As $\Delta P$ and $T_e$ are proportional, the P-Vref method tunes the PSS parameters such that the phase response of $G_C(s)$ is approximately zero. 
Specifically, the P-Vref method tunes the lead-lag block parameters such that the PSS TF phase response is the negative of the $G_{P\text{-}V\text{ref}}(s)$ phase response, resulting in 
$$\angle{G_C(s)}=\angle{G_{P\text{-}V\text{ref}} (s)} + \angle{F(s)} \approx0^\circ.$$

The phase response of $G_{P\text{-}V\text{ref}} (s)$ may be obtained by running simulations in which the AVR's reference voltage $V_\text{ref}$ is perturbed with a sinusoidal signal within the analyzed frequency range  from 0.1 to 10.0~Hz. The active power output of the corresponding SG is measured, and the phase shift between the input and output signals of $G_{P\text{-}V\text{ref}}(s)$ is calculated. This procedure is repeated for a number of frequencies within the desired range. During these simulations, the SG speed $\omega$ %(and the rotor angle $\delta$) 
should remain constant, as deviations in $\omega$ affect $T_e$ and $P$. This can be achieved by setting the inertia constants $H$ of all SGs in the power system model to a sufficiently high value, thus ensuring that the shaft dynamics are disabled, i.e., $\Delta \omega \approx0$.

The time constants of the lead/lag blocks $T_1$ to $T_4$ are tuned over the desired frequency range such that the phase response of the PSS best matches the negative of the calculated phase response of $G_{P-V\text{ref}} (s)$. 
The high-pass filter $T_\mathrm{W}$ is selected as done in the Residues method to target the frequencies of interest.
After the time constants are selected, the PSS gain $K_\text{PSS}$ is tuned using root locus analysis of $W(s)$ from \eqref{eq:residues_tf}, as with the Residues method.

%%%%%%%%%%%%%%%%%%%%%%%%%%%%%%%%%%%%%%%%%%%%%%%%%%%%%%%%%%%%%%%%%%%%%%%%%%%
\section{Results} \label{sec:results}
\subsection*{Case Study Setup}

%%%%%%%%%%%%%%%%%%%%%%%%%%%%%%%%%%%%%%%%%%%%%%%%%%%%%%%%%%%%%%%%%%%%%%%%%%%
\begin{figure}[b!]
    \centering
    \vspace{-0.6cm}
    \includegraphics[width=1.00\linewidth]{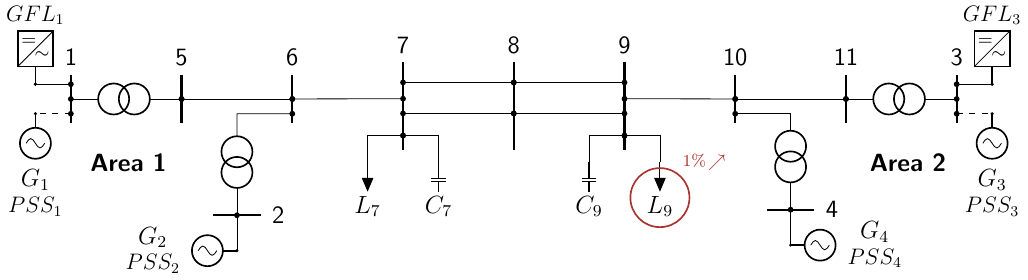}
    \vspace{-0.7cm}
    \caption{Two-area system used in simulations: 0\% IBR system (only SGs connected) and 50\% IBR system (SG at Buses 1 and 3 replaced with GFLs).}
    \label{fig:Kundur_System}
\end{figure}
%%%%%%%%%%%%%%%%%%%%%%%%%%%%%%%%%%%%%%%%%%%%%%%%%%%%%%%%%%%%%%%%%%%%%%%%%%%

%%%%%%%%%%%%%%%%%%%%%%%%%%%%%%%%%%%%%%%%%%%%%%%%%%%%%%%%%%%%%%%%%%%%%%%%%%%
\begin{figure*}[h!]
    \centering
    \includegraphics[width=\textwidth]{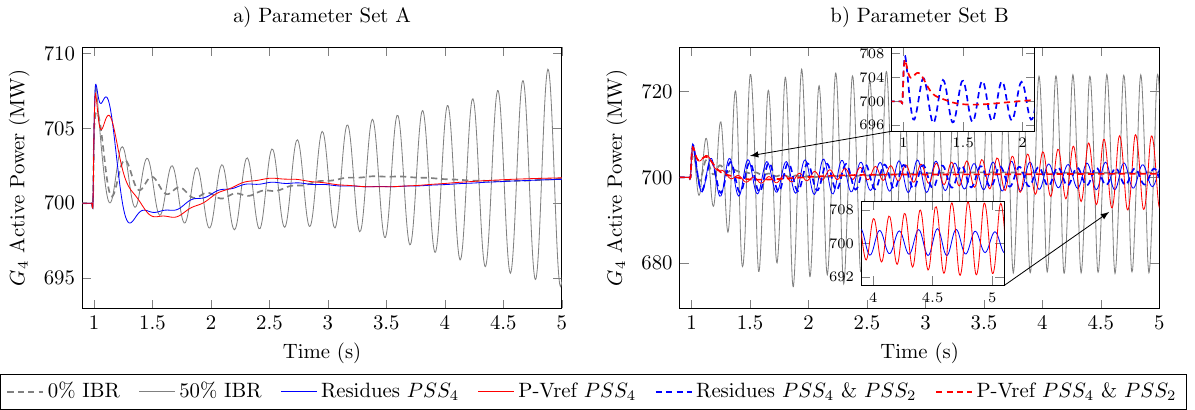}
    \vspace{-0.6cm}
    \caption{Comparison of SG $G_4$ active power output trajectories for the different analyzed parameter sets and PSS re-tuning methods.}
    \label{fig:results_trajectories}
    \vspace{-0.3cm}
\end{figure*}
%%%%%%%%%%%%%%%%%%%%%%%%%%%%%%%%%%%%%%%%%%%%%%%%%%%%%%%%%%%%%%%%%%

This section examines the influence of increasing IBR penetration on system stability and PSS tuning using dynamic simulations of the two-area system from \cite{Kundur1994} -- a benchmark known for exhibiting both inter-area (0.1--0.7 Hz) and local (0.7--2.0 Hz) electromechanical oscillations -- performed with the \textit{STEPSS} simulation platform \cite{RAMSES}. The system parameters can be found in \cite[Example~12.6]{Kundur1994}, corresponding to a thyristor exciter with high transient gain. The sixth-order SG model used in this study is from \cite{STEPSS2023}, while the speed governor represents a thermal turbine model with a reheater. The exciter and AVR model implemented across all generators is detailed in \cite{VanCutsem2013}, and the PSS model is described in Section~\ref{preliminaries:pss}. System loads are modeled as constant current for active power and constant impedance for reactive power.

The load $L_9$ in Area 2 is significantly larger than $L_7$ in Area 1, and the initial operating point shows a high power transfer from Area 1 (left) to Area 2 (right). This makes the system susceptible to poorly damped or even unstable inter-area oscillations. 

The GFL model used in this work is a standard GFL model. It includes a phase-locked loop for synchronization, an active power control loop that tracks the given setpoint, and a reactive power control loop that provides voltage support. The converter injects current into the grid through an $RL$ reactor. 

\subsection*{Case Study Comparisons}
We analyze two different dispatch scenarios of the system in Fig.~\ref{fig:Kundur_System}: a 0\% IBR scenario and a 50\% IBR scenario. The former consists of two identical areas (except for the loads), each comprising two SGs, one load, and a capacitor bank. In the latter system, the SGs (and their associated PSSs) at Buses 1 and 3 are replaced with GFL converters while maintaining the same generation and load balance. This replacement reduces system strength and makes the system more prone to oscillations as the PSSs have not been tuned for the new dynamics. 

The PSSs are installed at each SG and configured with identical parameters. We analyze two sets of PSS parameters, presented in Table~\ref{tab:results}, each of which could represent legacy-tuned PSS parameters:
\begin{itemize}
    \item \textit{Parameter Set A}: Residues-tuned PSS parameters for the 0\% IBR System under different loading conditions.
    \item \textit{Parameter Set B}: P-Vref-tuned PSS parameters for the 0\% IBR System under different loading conditions.
\end{itemize}
For both sets of parameters, the 0\% IBR System is stable. As will be shown, the two parameter sets behave differently for the 50\% IBR System.

To assess system stability, we disturb the system with a 1\% load increase at Bus 9 and observe the active power output trajectories of SG $G_4$ at Bus 4. Upon detecting instability, we re-tune the PSS at Bus 4 ($PSS_4$), identified by a participation factor and residue analysis as the most effective re-tuning location due to its dominant involvement in the unstable mode, without altering the remaining PSS at Bus 2 ($PSS_2$). If the system is still unstable after re-tuning $PSS_4$ (Section \ref{results:b}), $PSS_2$ is subsequently re-tuned.

%%%%%%%%%%%%%%%%%%%%%%%%%%%%%%%%%%%%%%%%%%%%%%%%%%%%%%%%%%%%%%%%%%
\begin{table}[b!]
    \centering
    \vspace{-0.5cm}
    \caption{Initial and re-tuned PSS parameters for test system in Fig.~\ref{fig:Kundur_System}.}
    \vspace{-0.7cm}
    \begin{center}
    \scalebox{0.823}{
    \begin{tabular}{ll|c|c|c|c|c|c|c}
    \toprule
    \multicolumn{2}{c|}{Parameters} & PSS & $K_\text{PSS}$ & $T_\mathrm{W}$ & $T_1$ & $T_2$ & $T_3$ & $T_4$ \\
    \midrule
    \multirow{2}{*}{\shortstack{Legacy\\tuned}} & Set A & \multirow{2}{*}{All} & 20 & 10 & 0.4863 & 0.1415 & 0.4863 & 0.1415 \\
    & Set B & & 40 & 10 & 0.2460 & 0.0352 & 0.2460 & 0.0352 \\
    \midrule
    \multirow{3}{*}{\shortstack{Residues\\re-tuned}} & Set A & $PSS_4$ & 34 & 10 & 0.2852 & 0.2162 & 1.0000 & 1.0000 \\
    & Set B & $PSS_4$ & 270 & 10 &  0.0426 & 0.0101 & 1.0000 & 1.0000 \\
    & Set B$^*$ & $PSS_2$ & 900 & 10 &  0.0629 & 0.0115 & 0.0629 & 0.0115 \\
    \midrule
    \multirow{3}{*}{\shortstack{P-Vref\\re-tuned}} & Set A & $PSS_4$ & 30 & 10 & 0.0407 & 0.0001 & 0.0407 & 0.0001 \\
    & Set B & $PSS_4$ & 45 & 10 & 0.0407 & 0.0001 & 0.0407 & 0.0001 \\
    & Set B$^*$ & $PSS_2$ & 80 & 10 & 0.0430 & 0.0001 & 0.0430 & 0.0001 \\
    \bottomrule
    \end{tabular}}
    \end{center}
    \label{tab:results}
\end{table}
%%%%%%%%%%%%%%%%%%%%%%%%%%%%%%%%%%%%%%%%%%%%%%%%%%%%%%%%%%%%%%%%%%

\subsection{PSS Parameter Set A} \label{results:a}
After transitioning from the base case to the 50\% IBR scenario, the system becomes unstable for a 1\% load increase at Bus 9, as indicated by the gray increasing-magnitude oscillations in Fig.~\ref{fig:results_trajectories}a. To re-tune $PSS_4$, we first deactivate it and then determine new PSS parameter values using either the Residues or P-Vref methods described in Section~\ref{sec:methodology}. We use the 50\% IBR scenario as the model for both PSS tuning methods. The $PSS_4$ re-tuning is carried out while keeping the remaining PSS, $PSS_2$, active. The re-tuned parameters of $PSS_4$ are summarized in Table~\ref{tab:results}. 

To test the stability of the system with a given PSS-tuning, we run tests in which the PSS parameters are set, and then the load at Bus 9 is increased at $t=1$~s. 

Fig.~\ref{fig:results_trajectories}a plots the active power trajectories of $G_4$ for each test. The simulations confirm the system is stable for both Residues (blue) and P-Vref-tuned parameters (red). Therefore, we conclude that, for Parameter Set $A$, tuning $PSS_4$ with either method is sufficient, and thus coordination is unnecessary. % The results for Parameter Set $B$ in the next section tell a different story, highlighting the sensitivity of power system stability to PSS tuning.

\subsection{PSS Parameter Set B} \label{results:b}

\subsubsection*{\textbf{$\boldsymbol{PSS_4}$ Re-tuning}} 
As with Parameter Set A, the growing oscillation magnitude in gray in Fig.~\ref{fig:results_trajectories}b indicates that the system is unstable after the disturbance for the 50\% IBR scenario. As described in Section~\ref{sec:results}, we begin the analysis by re-tuning just $PSS_4$ and then running the simulations with the load increase at $t=1$~s. The active power trajectories of $G_4$ with the re-tuned $PSS_4$ are shown by the solid lines in Fig.~\ref{fig:results_trajectories}b. The re-tuned parameters are presented in Table~\ref{tab:results} in the rows named Set B. 

\begin{figure*}[h!]
    \centering
    \vspace{-0.0cm}
    \includegraphics[width=1.00\linewidth]{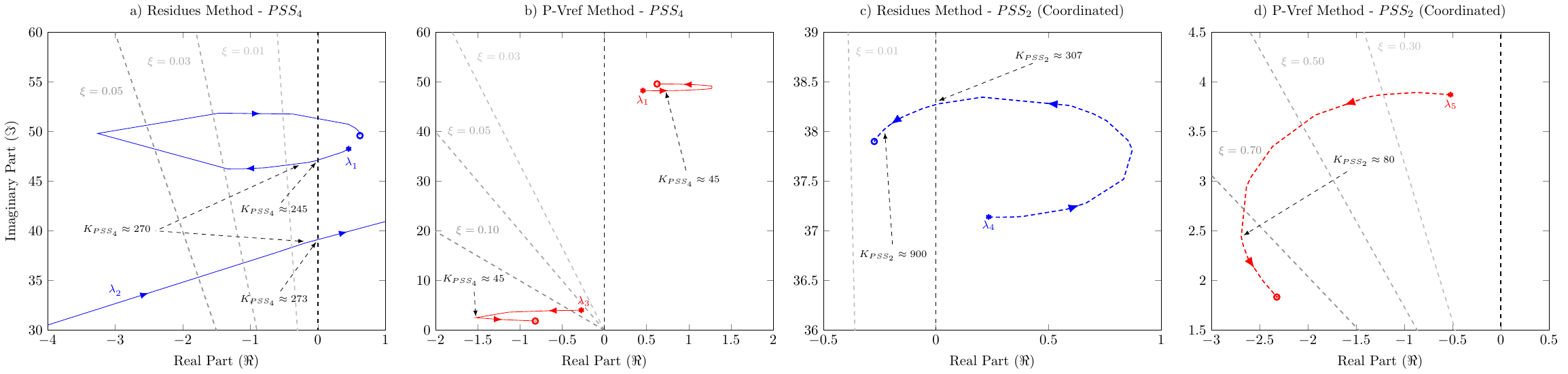}
    \vspace{-0.6cm}
    \caption{Movement of the critical poles $\lambda_j, j \in \{1, \dots, 5\}$ in the root-locus plots for the initial PSS Parameter Set B under different re-tuning scenarios: (a) and (b) show the re-tuning of just $PSS_4$; (c) and (d) illustrate the sequential re-tuning of $PSS_4$ and $PSS_2$. The gray dotted lines indicate the damping ratio of the modes $\xi=-\frac{\sigma}{\sqrt{\sigma^2+\omega^2}}$. The stars indicate the open-loop pole positions (with $K_\text{PSS}=0$), whereas circles mark the zeros towards which the closed-loop poles migrate along the plotted trajectories as the gain increases$^1$. The line styles correspond to those used in Fig.~\ref{fig:results_trajectories} for the respective PSS re-tuning scenario.}
    \vspace{-0.3cm}
\label{fig:results_modes}
\end{figure*}

The trajectories reveal that the system remains unstable after re-tuning $PSS_4$ using the P-Vref method, while the Residues-tuned $PSS_4$ exhibits poorly damped oscillations. The Residues-tuned system behavior can be attributed to the fact that the Residues method targets the unstable mode $\lambda_1$ (Fig.~\ref{fig:results_modes}a) and shifts it to the left-half plane. However, as the gain increases, the $\lambda_2$ mode approaches instability. As a result, all stable cases for the selected gain $K_{PSS_4}$ exhibit a small damping ratio $\xi$. On the other hand, the P-Vref method cannot shift the unstable mode $\lambda_1$ to the left-half plane with any gain, as shown in Fig.~\ref{fig:results_modes}b, due to the high influence of $PSS_2$ on~$\lambda_1$.

Based on this analysis, we conclude that for Parameter Set B and tuning just $PSS_4$, the Residues method outperforms P-Vref, but is still unsatisfactory.

\setcounter{footnote}{0}
%%%%%%%%%%%%%%%%%%%%%%%%%%%%%%%%%%%%%%%%%%%%%%%%%%%%%%%%%
%%%%%%%%%%%%%%%%%%%%%%%%%%%%%%%%%%%%%%%%%%%%%%%%%%%%%%%%%

\subsubsection*{\textbf{Uncoordinated PSS Re-tuning}} 

In uncoordinated PSS Re-tuning, each PSS re-tunes its PSS parameters simultaneously, without considering that the other PSS is also making adjustments.
For Parameter Set B, we observed that uncoordinated Residues PSS Re-tuning did not stabilize the grid. This is due to the ``wrong" $\mathbf{A}$ matrices used for tuning each PSS, which do not consider that the parameters of the other PSS will change. Uncoordinated P-Vref, on the other hand, did stabilize the grid and produced similar behavior to the Sequential P-Vref method described next, and plotted in Fig.~\ref{fig:results_trajectories}b with dashed red lines. The results are omitted here due to limited space.

\subsubsection*{\textbf{Sequential Coordinated PSS Re-tuning}}
Since re-tuning only $PSS_4$ fails to stabilize the system using the P-Vref method and yields unsatisfactory results with the Residues method, while uncoordinated PSS re-tuning works only for P-Vref but fails for Residues, we introduce a ``sequential'' tuning strategy: $PSS_4$ is re-tuned first, followed by $PSS_2$, with $PSS_4$ remaining active and re-tuned. We note that the sequential tuning strategy requires coordination between the two PSSs. The resulting trajectories for each re-tuning method are shown by the dashed blue (Residues) and red (P-Vref) lines in Fig.~\ref{fig:results_trajectories}b. Table~\ref{tab:results} lists the re-tuned parameters in Set B$^*$ rows.

The dashed trajectories in Fig.~\ref{fig:results_trajectories}b demonstrate that sequential tuning is effective with the P-Vref method, and does not change the performance of the Residues method significantly. The Residues method considers the full $\mathbf{A}$ matrix, and thus is not well-suited for sequential tuning---in this test case, the re-tuning of $PSS_4$ tries to stabilize the entire system, without recognizing that $PSS_2$ will subsequently be retuned and can help. That is, the Residues tuning of $PSS_4$ establishes new dynamics for which $PSS_2$ has little impact. As shown by the root-locus of $PSS_2$ re-tuning in Fig.~\ref{fig:results_modes}c, $PSS_2$ requires a large gain and still cannot push the pole far into the left-half plane.

In contrast, the P-Vref method is generator-specific, as it relies on the local phase response of the generator on which the PSS is installed. As a result, P-Vref tuning is more effective at targeting modes associated with that generator, rather than those linked to other machines that may also contribute to instability. Moreover, the local nature of the P-Vref method also makes it well suited for sequential tuning. This intuition is corroborated by Figs.~\ref{fig:results_trajectories}b~and~\ref{fig:results_modes}d, which confirm that sequential P-Vref tuning yields a well-damped and stable response.

We conclude that P-Vref is better suited for sequential re-tuning than Residues. 
The Residues method may be effective with other types of coordination than the simple sequential tuning approach investigated here, such as selectively turning off PSSs while tuning is conducted. However, such advanced coordination may not be viable in practice.
Furthermore, we note that the sequential P-Vref method requires coordination between the generators, which may still be challenging in practice for large systems.

%%%%%%%%%%%%%%%%%%%%%%%%%%%%%%%%%%%%%%%%%%%%%%%%%%%%%%%%%%%%%%%%%%%%%%%%%%%
\section{Conclusion} \label{sec:conclusion}
In this work, we analyze the performance of the standard Residues and P-Vref PSS tuning methods as GFL IBR penetration increases in the Kundur Two-Area System. The case study scenarios reveal that the effectiveness of the PSS tuning methods is not guaranteed in all circumstances and that, depending on the initial PSS parameters, coordination between PSSs may be necessary. If coordination is not possible, the Residues method fails to ensure small-signal stability, whereas the uncoordinated P-Vref method achieves a well-damped response. We note that, while tuning both PSSs with the P-Vref method worked in the case study we investigated, it is not guaranteed to work for all systems.

The findings in this case study motivate a local and adaptive PSS tuning method that can be run online, accommodating changing dynamics of the grid as IBR resources come online.

%%%%%%%%%%%%%%%%%%%%%%%%%%%%%%%%%%%%%%%%%%%%%%%%%%%%%%%%%%%%%%%%%%%%%%%%%%%
\newpage
\footnotetext[1]{The root locus plots arise after the Residues or P-Vref method determines the time constants $T_1$ to $T_4$. The open-loop poles correspond to turning off the given PSS. In Fig.~\ref{fig:results_modes}c, turning off $PSS_2$ destabilizes the system as the open-loop pole $\lambda_4$ is in the right-half plane. In Fig.~\ref{fig:results_modes}d, turning off $PSS_2$ stabilizes the system as all open-loop poles are in the left-half plane.}

% References section
\bibliographystyle{IEEEtran}
\bibliography{bibliography.bib}

\end{document}